\documentclass[sigconf,nonacm]{acmart}

\AtBeginDocument{%
  \providecommand\BibTeX{{%
    \normalfont B\kern-0.5em{\scshape i\kern-0.25em b}\kern-0.8em\TeX}}}

\begin{document}

\title[Assessing the Quality of Computational Notebooks]{Assessing the Quality of Computational Notebooks\\for a Frictionless Transition from Exploration to Production}

\author{Luigi Quaranta}
\email{luigi.quaranta@uniba.it}
\orcid{0000-0002-9221-0739}
\affiliation{%
  \institution{University of Bari}
  \streetaddress{Via Edoardo Orabona, 4}
  \city{Bari}
  \country{Italy}
  \postcode{70125}
}

\begin{abstract}
  The massive trend of integrating data-driven AI capabilities into traditional software systems is rising new intriguing challenges. One of such challenges is achieving a smooth transition from the explorative phase of Machine Learning projects -- in which data scientists build prototypical models in the lab -- to their production phase -- in which software engineers translate prototypes into production-ready AI components. To narrow down the gap between these two phases, tools and practices adopted by data scientists might be improved by incorporating consolidated software engineering solutions. In particular, computational notebooks have a prominent role in determining the quality of data science prototypes. In my research project, I address this challenge by studying the best practices for collaboration with computational notebooks and proposing proof-of-concept tools to foster guidelines compliance.
\end{abstract}


\begin{teaserfigure}
  \vspace{0.5cm}
\end{teaserfigure}

\maketitle

\section{Research Problem}
\label{sec:research-problem}

Over the past few years, Artificial Intelligence (AI) and Machine Learning (ML) have seen unprecedented development, thus enabling the design of intelligent software solutions in a large and growing number of application domains.
As a direct consequence, the integration of AI-powered capabilities in traditional software systems has become a massive trend \cite{amershi_software_2019}.

Such an integration, however, poses a large number of complex challenges \cite{kim_data_2018, arpteg2018software, lwakatare_taxonomy_2019, nascimento_understanding_2019, lwakatare_large-scale_2020}, in part inherited from the fields of Software Engineering (SE) and Data Science (DS), in part completely new \cite{wan_how_2019}, arising from the need to translate lab model prototypes into production-ready AI/ML components \cite{lanubile_towards_2021, lewis_component_2019, lewis_characterizing_2021}.

Despite the availability of high-level libraries and frameworks 
that make the development of AI/ML prototypes a relatively easy task to achieve -- even for professionals lacking a computer science background --
building reliable and scalable AI systems remains a complex endeavor.
Uninformed misuses of low-code ML frameworks typically end up being the source of new kinds of technical debt \cite{sculley_hidden_2015, bogner_characterizing_2021, tang2021empirical}.

A possible way to approach the multifaceted challenge of developing production-ready AI components is to apply -- early in the ML model building process -- suitable adaptations of consolidated Software Engineering solutions \cite{washizaki_studying_2019, serban_adoption_2020, van_oort_prevalence_2021}.
For instance, practitioners have already started leveraging Continuous Integration (CI) and Delivery (CD) in ML \cite{sato_continuous_2019}.
As a doctoral student, I have embraced this perspective and tried to address -- through the lens of a software engineer -- the research problem of achieving a smoother translation of AI/ML prototypes to production.

So far, the focus of my research has been on the workflow and tooling employed by DS teams in the early stages of AI/ML projects, with a particular emphasis on one of the most popular tools among data scientists: the computational notebook \cite{rule_exploration_2018}.

\section{Hypothesis}
\label{sec:hypothesis}

\begin{figure*}[ht]
  \centering
  \includegraphics[width=0.56\textwidth]{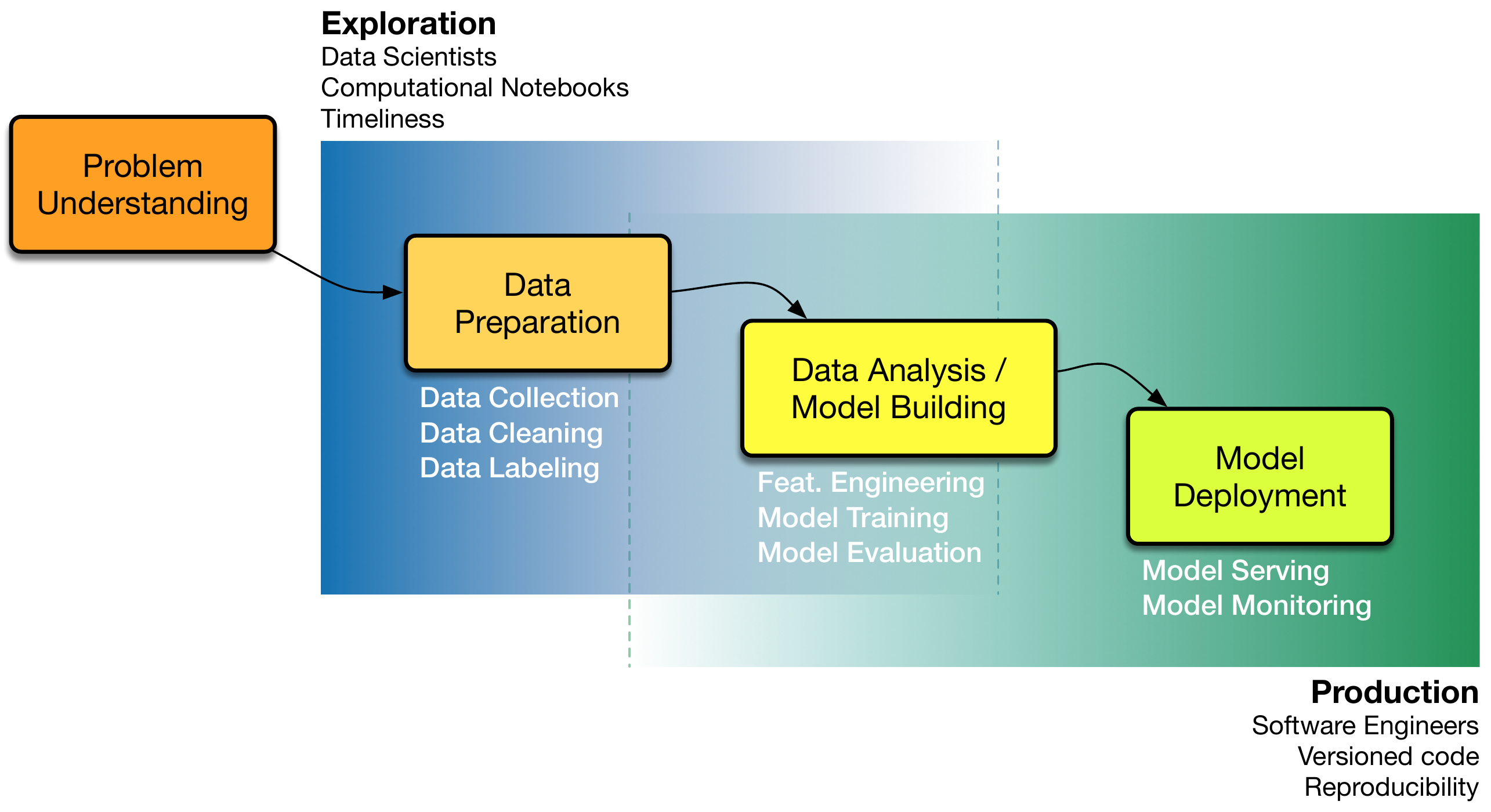}
  \caption{Tension between \textit{exploration} and \textit{production} in the typical ML workflow.}
  \label{fig:workflow}
  \vspace{-2.6mm}
\end{figure*}

Having had the opportunity to complement a preliminary literature review with qualitative research conducted at an Italian Data Science company, early in my PhD studies I have been able to detect and characterize a tension between the initial, exploitative phase of an AI/ML project and the final, production phase. This tension, mainly caused by a shift in the goals of the two phases, is exacerbated by the different educational backgrounds of the stakeholders involved and the different tools in use.
This idea is illustrated in Figure~\ref{fig:workflow}, depicting a typical ML workflow.

In the initial phase of an AI/ML project -- which we will call the \textit{Exploration} phase --  data scientists generally build prototypical models using computational notebooks, i.e., interactive documents combining code and natural language text in a cohesive narrative of experimental trials. Their main goal is to timely deliver the best-performing models to the rest of their team. 

Later, in the final phase of a project -- hereafter referred to as the \textit{Production} phase -- the best ML prototypes are handed off to software engineers, who are typically in charge of bringing them to production as live services; to achieve this, software engineers leverage consolidated SE solutions (e.g., VCS, testing frameworks, static code analyzers, etc.) and practices (e.g., agile practices, CI/CD). Their goal is to build robust and scalable AI components out of ML prototypes and to consolidate the resulting process into a reproducible pipeline.

Transitioning from \textit{Exploration} to \textit{Production} is not easy, even for companies that rely on well-established teams and procedures. Typically, the two phases have no clear boundaries: it is not easy to define when, in an ML workflow like the one in Figure~\ref{fig:workflow}, the transition should happen and who is accountable for it: should data scientists be in charge of raising the quality of their code to the production standards? Or should software engineers be accountable for organizing and improving prototypical code, making it production-ready?

My research work is based on the following assumption: incorporating software engineering solutions into the tools and practices adopted by data scientists in the early stages of AI/ML projects can smooth out the friction that is currently experienced by practitioners in the transition between the aforementioned two phases and reduce the ``cultural'' mismatch between DS and SE professionals.
Furthermore, given the widespread adoption of computational notebooks within the data science community, I hypothesize that engineering-aware enhancements of popular notebook platforms like Jupyter Notebook \cite{perez_project_2015} would produce the most tangible effects on the production-readiness of prototypical ML artifacts.

\section{Expected Contributions}
\label{sec:expected-contributions}

With my PhD research work, I expect to contribute to the development of better, SE-aware collaborative practices in data science teams, with a focus on the use of computational notebooks.

Collaboration in data science teams happens across a wide range of tools and practices \cite{zhang_how_2020}. Among these, computational notebooks -- and in particular Jupyter Notebook -- have seen a steadily growing adoption by data scientists worldwide as they offer a fast and interactive prototyping environment, a lightweight form of experiment documentation, and innovative collaboration capabilities, powered by modern web technologies~\cite{perkel2018jupyter}. Nonetheless, the notebook format has also been criticized for inducing bad programming practices (e.g., code fragmentation and non-linear execution of code) as well as for offering scarce-to-null native support for Software Engineering best practices (even basic ones, such as code modularization, testing, and versioning) \cite{grus_i_2018}. Consequently, many researchers have started investigating the quality of Jupyter notebooks and proved that they are too often (\textit{i}) inundated by poor-quality code \cite{wang_better_2020, pimentel_large-scale_2019, koenzen_code_2020}, (\textit{ii}) scarcely reproducible \cite{pimentel_large-scale_2019}, and (\textit{iii}) superficially documented \cite{rule_exploration_2018} -- notwithstanding their support for richly formatted text and multimedia output.

To counter these phenomena, some scholars have already proposed best practices for better use of computational notebooks \cite{rule_ten_2018, pimentel_large-scale_2019}, although not providing scientific validation for them. Building on their work, I have already contributed by collecting and validating a comprehensive list of best practices for the collaborative use of computational notebooks in a professional context.

Other researchers have built their own extensions for the Jupyter ecosystem, aimed at supporting the creation of better notebooks. The contributions range from notebook-specific versioning systems \cite{kery2018interactions}, to tools that improve the readability \cite{rule_aiding_2018} and the communication capabilities of notebooks \cite{wang_callisto_2020}. Head et al. also proposed an extension for the extraction of clean code from notebook cells, based on the desired output \cite{head_managing_2019}. Adding on these efforts, I plan to contribute by (\textit{i}) reviewing the currently available software solutions for the reproducibility of AI/ML experiments -- with an emphasis on notebook support -- and (\textit{ii}) designing, developing, and validating a linter for Jupyter Notebook, aimed at fostering the adoption of validated best practices for the collaborative use of this tool.

Ultimately, I will further investigate the problem of notebook reproducibility, in an original attempt to link it with the adoption/violation of best practices.

\section{Evaluation Plan}
\label{sec:evaluation-plan}

As stated in the previous section, one of the major expected contributions of my research is the collection and validation of a catalog of best practices for the collaborative use of computational notebooks. At present, I have already collected the best practices by means of a multivocal literature review (see \textsc{Study~3} in Sect.~\ref{sec:cscw2021}). For their validation, I followed a mixed approach: first, I qualitatively assessed the appropriateness and completeness of the catalog by conducting interviews with expert data scientists; then, I quantitatively assessed the adoption of the best practices by analyzing a selection of notebooks collaboratively developed by experts.

As for the second major contribution of my research, i.e., the development of a linter for Jupyter Notebook documents (see \textsc{Study~5} in Sect.~\ref{sec:pynblint}), I plan to validate it with a field study, to be performed in a professional context with professional data science teams.

Finally, to investigate the relationship between the reproducibility of computational notebooks and the adoption of best practices for their development, I plan to perform a further archival study (see \textsc{Study~6} in Sect.~\ref{sec:final-reproducibility-study}). This time, by leveraging the validated linter, I will seek for best practice violations, while also trying to reproduce the assessed notebooks.

\section{Preliminary Results}
\label{sec:preliminary-results}

In the following, I present the results achieved so far in my research work, going through the studies completed in the last two years.

\subsection{\textsc{Study 1}: Bringing ML models to production: an industry perspective from data scientists}
\label{sec:wain2021}

I started my research project by studying the challenges that lie in the integration of data-driven AI components into traditional software systems.

Upon completing a preliminary literature review on the subject, just a few weeks before the pandemic outbreak, my research group had the chance to get in touch with an Italian consulting company focused on data science, operating in the financial domain.
There, we organized a workshop on the reproducibility of AI/ML experiments.
The workshop was attended by 18 data scientists with varied educational backgrounds.
After an ice-breaking presentation on the existing software solutions for achieving reproducibility in AI/ML experiments, the event culminated in an inspiring brainstorming session on the topic of ML models productization. After the workshop, I performed a thematic analysis of the transcripts from the event, extracting the most relevant themes. 

One of the topics that soon catalyzed the debate was \textit{the role of computational notebooks in the typical ML workflow}. Some of the attendees deemed notebooks as indispensable tools, mainly because of their storytelling capability; when a data scientist needs to reconstruct the history and reasoning behind an experiment, having a well-written computational narrative is -- in practice -- more useful than VCS or other tracking mechanisms. Others promoted the use of computational notebooks alongside traditional IDEs: consolidated code fragments should be regularly transferred from notebook cells to a versioned and structured codebase. On the other end of the spectrum, a few attendees claimed that notebooks should be dropped as soon as possible during the workflow, immediately after the data exploration stage. Interestingly though, none of the attendees questioned the importance of computational notebooks: the discussion dealt, instead, with the right time in the workflow in which notebooks should be dismissed to give way to standard code.

Besides discussing the notebook lifecycle, some of the data scientists also expressed the \textit{desire for specific notebook features}: e.g., the native support for testing and versioning (since the data exploration stage), along with concrete aids to improve the reproducibility of computations within the Jupyter ecosystem.
    
Lastly, some of the attendees underlined the \textit{importance of defining a set of validated best practices}, to be shared across the team, and of fostering knowledge-sharing opportunities, to smooth out the cultural differences between data scientists and software engineers.

This study was accepted and presented at WAIN 2021~\cite{lanubile_towards_2021}. A few weeks after the workshop, I also administered a couple of surveys and performed some follow-up interviews with the members of the data science team of the company, with the specific aim of investigating their typical workflow. Overall, the insights gained in this phase constitute the foundation of the research hypothesis stated in Sect.~\ref{sec:hypothesis} and have inspired the focus of the rest of my work.

\subsection{\textsc{Study 2}: Software solutions for the reproducibility of AI/ML experiments}
\label{sec:aixia2021}

To gain a better understanding of the theme of reproducibility in data-driven AI experiments, I performed an exploratory review of the available software solutions that support its achievement. I tested each of the examined solutions against a predefined case study, taking notes about specific tool capabilities and their impact on the workflow. From this practical experience, I derived a taxonomy\footnote{Available at \url{https://github.com/collab-uniba/Software-Solutions-for-Reproducible-ML-Experiments}.} that can be leveraged by practitioners to orient themselves in the vast and growing offer of tools in this category. The details about the taxonomy are reported in an article that was accepted and will be presented at AIxIA 2021~\cite{quaranta_taxonomy_2021}.

\subsection{\textsc{Study 3}: Best practices for collaboration with computational notebooks}
\label{sec:cscw2021}

Inspired by the results of the workshop described in Sect.~\ref{sec:wain2021}, I started focusing my attention on computational notebooks. In particular, to meet the need for a set of validated best practices, I investigated the following research questions: (\textbf{RQ1}) \textit{``What are the best practices for collaboration with computational notebooks?''} and (\textbf{RQ2}) \textit{``What collaboration-specific best practices do experienced data scientists actually follow when working with computational notebooks?''}

To answer RQ1, I conducted a multivocal literature review \cite{garousi_guidelines_2019}, i.e., a systematic review in which sources can belong to both white literature (e.g., peer-reviewed papers published in journals or conference proceedings) or grey literature (e.g., informal documents produced by professionals, as technical blog posts). Upon performing a thematic analysis of the retrieved articles, I was able to collect a catalog of 17 best practices, organized in 6 main themes, for the collaborative use of computational notebooks.

To answer RQ5, I performed a qualitative and quantitative assessment of the adoption of the best practices from the catalog. The former was based on semi-structured interviews with 22 experienced data scientists from different companies; all interviewees had at least one year of experience with Jupyter Notebook. At the beginning of each interview, to provide some context, I broadly introduced the main themes that emerged in the multivocal literature review, although not revealing any specific guideline. Then, I asked participants to share their personal best practices for the collaborative use of computational notebooks. 

As for the quantitative assessment, it was performed by means of an archival study on a dataset of 1,380 Python Juyter notebooks taken from \url{kaggle.com}. Kaggle is a Google-owned platform hosting ML competitions for data scientists of all levels of expertise and offering a cloud-based data science environment in which users can write scripts and notebooks in the Jupyter format. A daily dump of the platform metadata is available as a Kaggle dataset named ``Meta Kaggle\footnote{\url{https://www.kaggle.com/kaggle/meta-kaggle}}.'' From this dataset, I assembled the collection of notebooks used in this study by filtering and downloading notebooks that were collaboratively authored by expert Kaggle users.

Overall, I found that practitioners are generally aware of the best practices available in the literature, although they do not always apply them. Indeed, the recommended behaviors are often deemed unfeasible or even counterproductive in definite contexts, mainly due to the lack of time and proper tool support. By reasoning on these results, I was able to identify possible enhancements for notebook development environments like Jupyter Notebook. This study is currently under minor revision at CSCW 2022~\cite{quaranta_title_2022}.

\subsection{\textsc{Study 4}: A dataset of Python Jupyter notebooks from Kaggle}
\label{sec:msr2021}

Having personally experienced the effort of putting together a dataset of computational notebooks with rich metadata for my archival studies, I decided to leverage the familiarity gained with the Kaggle platform to build \textsc{KGTorrent}, a dataset of Python Jupyter notebooks that I made publicly available to the research community.

\textsc{KGTorrent} consists of two main components: (1) the actual corpus -- comprising 248,761 Python Jupyter notebooks -- and (2) a companion MySQL database, derived from Meta Kaggle. Besides notebook metadata, the companion database stores comprehensive information about Kaggle users and competitions. 
The scripts developed to assemble \textsc{KGTorrent}, available on GitHub,\footnote{ \url{https://github.com/collab-uniba/KGTorrent}} can be used to replicate the collection as well as to update it according to newer versions of Meta Kaggle.

The paper on \textsc{KGTorrent} was accepted and presented at MSR 2021~\cite{quaranta_kgtorrent_2021}.
At present, I am also finalizing the development of an RPC-style web API for \textsc{KGTorrent}, to provide interested researchers with means to easily access the collection and filter it according to their specific research questions.

\section{Completion Plan}
\label{sec:completion-plan}

As clarified next, I plan to complete my research project with the development and validation of a proof-of-concept tool, aimed at improving the collaborative practices around computational notebooks, with the ultimate goal to smooth out the transition from exploration to production in AI/ML projects (see \textsc{Study~5} in Sect.~\ref{sec:pynblint}).
Moreover, I plan to perform an archival study to assess the impact of the adoption of best practices on notebook reproducibility (see \textsc{Study~6} in Sect.~\ref{sec:final-reproducibility-study}).

\subsection{\textsc{Study 5}: pynblint, a linter for Python Jupyter notebooks}
\label{sec:pynblint}

At present, I am designing and developing a linting library for computational notebooks, named \texttt{pynblint}. Specifically, the library will perform the analysis of data science projects containing Python Jupyter notebooks: each notebook in a project repository will be checked against a predefined set of rules. Such rules will be based on operationalizations of the best practices from the catalog introduced in Study~3 (see Sect.~\ref{sec:cscw2021}): \texttt{pynblint} will warn its users when they do not comply with the guidelines and promptly suggest corrective actions.
Julynter, a similar contribution by Pimentel et al.~\cite{pimentel_understanding_2021}, is implemented as a plugin for a specific notebook platform: JupyterLab.
Differently, \texttt{pynblint} will be a static code analyzer for all notebooks in the Jupyter format, regardless of the platform used to write them. 
It will work both as a standard CLI application -- to be seamlessly integrated into CI/CD pipelines -- and as a modern web app -- to be easily accessed by beginners. I plan to conclude this activity by February 2022.

In the following three months, after a pilot test of the tool with student volunteers from a data science course of my university, I plan to conduct a thorough validation of \texttt{pynblint} in the field, by involving one or more professional data science teams.
The expected output of this validation process is two-fold: (\textit{i}), the collected feedback will enable a refinement of the linting library; (\textit{ii}) I will be able to gather further statistics on the adoption of the guidelines, this time enriched with contextual information (i.e., the educational background of the notebook author, the application domain of the project, the time-span allotted to the project, etc.); consequently, I will be enabled to determine, in each context, which of the proposed fixes are typically adopted and which ignored.

\subsection{\textsc{Study 6}: Sharing reproducible notebooks -- are best practices enough?} 
\label{sec:final-reproducibility-study}
Non-reproducible notebooks hinder collaboration and hamper the transition from exploration to production.
To address this further concern, I will conclude my research project by investigating the following research questions: (\textbf{RQ3}) \textit{``Is there a relationship between the reproducibility of computational notebooks and the adoption of the best practices for their collaborative development?''} and (\textbf{RQ4}) \textit{``What are the guidelines that are most beneficial to the reproducibility of computational notebooks?''}. I plan to accomplish this investigation by the end of June 2022.

To answer the questions, upon completion of the \texttt{pynblint} validation process, I will leverage the linter and build a record of the guideline violations found in a dataset of Python Jupyter notebooks, while also testing the reproducibility of the related computations. Then, I will be able to statistically determine the existence of a relationship between the adoption of the best practices and the reproducibility of computational notebooks.

\section{Conclusion}

In my research project, I address the challenging translation of AI/ML prototypes into production-ready AI components. In particular, given their widespread adoption among data scientists, I have investigated the best practices for the collaborative use of computational notebooks. 
Furthermore, in the last year of my PhD, I will conclude the development and validation of a dedicated linter to check and foster compliance of data science repositories with the guidelines.
I plan to defend my doctoral thesis in early 2023.

\bibliographystyle{ACM-Reference-Format}
\bibliography{references}

\end{document}